# Introducing the "comfort performance gap" in new educational buildings – a case study

Theresa Kohl[1], Thomas Schranz[1], Eva Hofmann[2], Katja Corcoran[2], Gerald Schweiger*[1]

[1] Graz University of Technology, Institute of Software Technology, 8010 Graz, Austria

[2] University of Graz, Institute of Psychology, 8010 Graz, Austria

*gerald.schweiger@tugraz.at

## ABSTRACT

Providing adequate indoor environmental quality is crucial in educational settings. In this paper, we implemented and tested a framework that collects occupant feedback and investigated correlations between teachers' comfort and the operational characteristics of an Austrian school building in September and October 2022. Initial results show that the measured average temperatures (23.1-25.1°C) in all rooms are on the upper limit of various recommendations, such as comfort guidelines for building operation or workplace regulations. This assessment is in line with the feedback we received from the teachers. A literature review demonstrated that children's comfort temperatures are lower compared to adults. Hence, it is reasonable to conclude that indoor temperatures during the survey period were inadequate for the pupils either, even without direct feedback. An analysis of the CO2 measurements showed that, during school hours, approximately 20% of all measurement values were above 1000 ppm, with 2% above 1500 ppm. CO2 levels above 1000 ppm are considered hygienically critical, with the latest research proposing to lower the limits below 800 ppm to ensure a healthy and effective learning environment. While we only assessed the challenges of providing a healthy indoor environment for an educational building in Austria, our literature review shows similar challenges and research efforts worldwide. Our analysis demonstrates the need for adapting design requirements, especially for school buildings, acknowledging the different comfort needs of adults and children and the importance of high indoor air quality for providing an optimum learning environment. Future research should focus on testing adapted indoor environmental quality requirements for schools, especially in urban areas, and how to integrate real-time occupant feedback in the heating, ventilation and air conditioning systems.

## INTRODUCTION

The European Green Deal provides the framework to achieve climate neutrality by 2050 in the European Union [1]. Today, buildings use 40% of the European energy demand, resulting in a share of 36% of greenhouse gas emissions [2]. Over the past 20 years, the EU established a legislative framework around the European Energy Performance of Buildings Directive (EPBD) to increase energy efficiency. Within this framework, the member states implemented the “Nearly Zero Energy Building Standard” by which, after December 2018, all new public buildings, and by December 2020, all new buildings must be nearly zero energy. Besides improving energy efficiency, member states have to ensure adequate indoor climate [3]. The new EPBD recast, proposed in December 2021, defines even stricter energy performance requirements (“Zero-Emission Building Standard”) and emphasizes the importance of maintaining healthy indoor air quality. The recast requires equipping zero-emission buildings with sensors and control devices to monitor and regulate indoor air quality. In addition, the recast requires regular inspection of ventilation systems with an effective output greater than 290 kW to ensure their reliability [4].

Since approximately 85% of the European educational buildings were built before the year 2000 [5], and the first version of the EPBD entered into force in 2002, there is a tremendous need for renovation activities to increase energy performance and user comfort. Although newer buildings already use advanced technologies, such as building automation for heating, cooling, ventilation (HVAC), lightning, or shading, they are not well adapted for building renovations [6]. However, detailed data about the technical equipment of the European building stock is rare and only available for some countries [7]. As an example, in the Swiss building stock, very few educational buildings built before 2009 are equipped with a ventilation system (5-30%) and/or a cooling system (1-10%) [8]. Therefore, it is reasonable to assume that in most educational buildings, it is impossible to monitor or control healthy indoor air quality actively.

Measuring whether an indoor environment is healthy is often linked to assessing the indoor environmental quality (IEQ) parameters of a building, determined by indoor air quality (IAQ), thermal comfort, lighting, and acoustics [9]. IAQ includes indoor temperatures, humidity, gaseous pollutants, and aerosols [10]. According to Tappler (2017), CO2 is commonly used as the main overall indicator for IAQ, although it is not the only criterion for evaluating indoor climate [11].

**Background**

People spend up to 90% of their time indoors [13]. Consequently, the IEQ significantly impacts the health and well-being of building occupants [12, 13]. Especially respiratory diseases such as asthma are caused by poor IAQ, and occupants often suffer from what has been termed “sick building syndrome” [14]. Recently, the COVID-19 pandemic demonstrated that well-functioning ventilation in buildings is one of the most effective methods to remove virus-containing aerosols, serving as a precaution against infection [15, 16]. Additionally, IEQ is crucial in educational settings. Studies found that low IEQ decreases comfort and harms children’s attendance and performance [17, 18, 19, 20] as well as teachers’ productivity [21]. Shendell et al. (2004) even found a statistically significant association between CO2 levels and absenteeism in schools [22].

Dorizas et al. (2018) summarized different investigations about the effects of building design and operation on productivity and indicated that adequate lighting improves learning progress by up to 26% as well as reading speed and concentration performance by up to 30% [9]. Julios et al. (2007) provided evidence that asthma-related hospital admissions of children peak at the beginning of the new school year in September in two cities in England and Scotland [23]. Simoni et al. (2010) found significant relations between respiratory disturbances and nasal patency in poorly ventilated classrooms [24].

Perceived thermal comfort depends on factors such as age, gender, physical activity, clothing, and state of health [26]. However, many standards for thermal comfort, such as the ISO 7730:2005 or the ASHRAE 55-2020, were designed under laboratory conditions and aimed at the average perception of thermal comfort of an entire group through measures such as the Predicted Mean Vote (PMV) or the predicted percentage of dissatisfied (PPD) [25]. Design requirements, such as thermal criteria, air quality, and ventilation, lighting, acoustics, aimed at minimizing the level of occupant dissatisfaction are commonly used within the industry [9]. However, it is difficult to integrate aggregated, static models such as PMV into a building control system. These models typically require input variables that are difficult to measure, such as clothing factors or metabolic rates, and are very imprecise for comfort predictions even if all the necessary variables are available under real operating conditions [26, 27, 28].

Besides standards for thermal comfort, other guidelines, such as design standards for HVAC systems, define room temperature ranges. Note that all these guidelines directly affect building energy consumption. Table 1 provides an overview of the applicable Austrian framework. Similar standards and recommendations for other European countries are widely available [9]. Although the EPBD and EPBD recast highlight the relation between a healthy indoor climate and energy efficiency, the European Union does not provide a harmonized framework with dedicated indoor climate indicators and corresponding thresholds, nor is indoor air quality reflected in the long-term renovation strategies of the member states. Fabbri & Dorizas (2019) provided comprehensive recommendations for addressing this issue in the relevant European directives [29].

We can categorize indicators for temperature and CO2 levels, such as the ones provided in Table 1, into design criteria used for planning and simulation and operational indicators for continuous monitoring. Although there is a plethora of guidelines from both categories, building operators lack comprehensive quality management tools to ensure the smooth and efficient operation of technical building facilities [36]. As a result, several studies and literature reviews in the past years analyzed the deviation between the predicted and actual energy performance of a building [37, 38, 39]. For instance, Menezes et al. (2012) reported gaps between predicted and measured energy consumption of 250% [40].

Besides the energy performance gap, buildings also fail to regularly achieve healthy IEQ and IAQ requirements. As previously hinted, this is particularly relevant in schools. Vornanen-Winqvist et al. (2020) conducted a comprehensive IAQ study in six public schools in Helsinki and found that IAQ complaints are common, even in the new buildings [41]. Cai et al. (2021) investigated IAQ, including particulate matter concentrations, in 33 classrooms in 21 schools with mechanical and natural ventilation in heavily polluted areas of Beijing. In classrooms with mechanical ventilation, the mean indoor temperature was 23,1°C (standard deviation 2,8°C), and the mean $CO_2$ level was 953 ppm (standard deviation 278 ppm) during the transitional season. $CO_2$ concentrations above 1000 ppm were measured 40% of the time when classes were occupied. [42]

Evidence suggests that children perceive thermal comfort differently than adults. In their literature review of comfort studies in educational settings of the past 50 years, Singh et al. (2019) showed that students in all age groups prefer cooler learning environments than adults [43]. Based on an investigation of the thermal performance of four school buildings, Teli et al. (2017) found growing evidence for children having lower comfort temperatures. With temperature thresholds adapted to children's preferences, they found that approximately 45% of the classrooms were overheated, compared to 0% if the adult-based comfort model was applied [44]. Nam et al. (2015) investigated seasonal differences in the comfort temperatures of children and adults, finding that children aged 4-6 years preferred lower temperatures (3.3°C in winter and 0.5°C in summer times) [45]. Based on investigations in eight primary schools in England, Bakó-Biró et al. (2012) proposed to keep temperatures between 20-22°C in winter and 22-24°C in summer to ensure an optimal learning environment [20]. However, additional investigations are required to predict the thermal comfort perception among school-aged children, contingent upon the prevailing thermal circumstances within classrooms [43]. The reasons for different thermal perceptions of adults and children are manifold. Possible explanations for why children perceive thermal comfort differently include differences in metabolic rates caused by physical exercise, lower capacities for acclimatization [45], their domestic environment, and outdoor play [43]. Empirical evidence shows that social background and behavior influence thermal preferences [46]. Additionally, children are assumed to have different thermoregulatory mechanisms because of the high body surface-to-mass ratio and limited sweating capabilities [46].

Occupancy density (persons per m²) also influences IAQ in schools. Several studies [47, 48, 49] showed that occupancy density in educational settings is 3-4 times higher than in offices. Note that occupancy directly affects required ventilation rates as well as heating and cooling loads.

Zhang et al. (2023) claim their findings confirm the necessity of addressing IAQ problems in schools worldwide [50]. Their research indicates that standard-compliant CO2 levels, air temperatures, and ventilation rates are not achievable in existing school classrooms due to design insufficiencies, affecting both naturally and mechanically ventilated classrooms. Singh et al. found similar results [43]. They evaluated 93 studies on thermal comfort in classrooms from the past 50 years, concluding that existing design guidelines and comfort standards are insufficient. Zou et al. (2018) showed that occupants and building facility managers are considered the most critical stakeholders in the operational phase of a building. However, facility managers struggle to implement efficient control strategies that serve occupants' needs and satisfy energy efficiency requirements [39]. According to Plesser (2020), implementing quality assurance during construction, adjustment of technical installations, and operation requires tremendous effort.

It stands to reason that facility management needs an efficient way to collect and manage large amounts of data [36]. Emerging technologies such as Building Information Modelling (BIM), the Internet of Things (IoT), and machine learning-based services are promising solutions to overcome data scarcity and obscure data management [51]. Tang et al. (2019) believe that by complementing BIM with IoT devices, it is possible to visualize and manage operational data, monitor IEQ, enhance user comfort, and utilize semantic web technologies to monitor and evaluate building performance in real time. Based on 97 reviewed articles, the authors proposed five integration methods. They highlighted a hybrid approach using semantic web and relational databases to store cross-domain data as the most promising, as it allows retaining different data sources in their platforms while achieving interlinking [52].

Besides effective data management, new technologies enabling bidirectional communication encourage occupants to actively participate in the operation of a building [53]. For instance, Schranz et al. (2020) developed a gamified, user-centered energy monitoring software, which allowed them to include user feedback in load predictions [54]. Bidirectional communication in real-time enables innovative user comfort paradigms, such as "Personal Comfort Models", which are currently being explored by various researchers [25]. Advances in the field of user-centered building services and automated building control have motivated research into the development of Human-in-the-Loop controls, such as dynamic temperature setpoint adjustments in buildings [67], or the derivation of "Cohort Comfort Models" from collected data [68].

### Main contribution

This work is a pre-study where we test a custom-developed web app-based data collection pipeline in a newly built school with reported IAQ complaints. We asked 22 teachers to share their perceptions of the indoor climate in September and October 2022 and correlated them to indoor climate data retrieved from the building management system (BMS) and its related sensing network. In the upcoming sections, we address the following research questions:

- What are the levels of $CO_2$ concentration and temperature in the classrooms during the survey period?
- How is the comfort of the teachers? How do they perceive performance of the pupils? Are there any relations between the feedback and the measurement data?
- What influence factors on comfort ratings were identified?
- How suitable is the existing building equipment in fulfilling the user's and building operator's needs?
- What are the reasons for the “comfort performance gap”?

The findings serve as a basis for future research activities, such as Human-in-the-loop control strategies and innovative comfort models. The remainder of the paper is divided into four sections and explores the answers to the research questions above. Section two gives a short overview of the background and related work in building technologies, comfort studies in schools, and the legal framework and describes the case study school. Section three provides details about data collection and data analysis. In section four, we discuss the results of the evaluation. The last two sections present a discussion of the limitations of our study and a conclusion.

## METHOD

The comfort study presented in this work was carried out from 12.09.2022-28.10.2022 and is based on two data sources: (i) operational data from the building, including room temperatures and CO2 concentration, (ii) feedback from 22 teachers, collected on an hourly and daily scale through a web application. Subsequently, we describe the case study, the relevant legal framework, and the data collection process.

### Case study

The investigated primary school building (4256 m²) is situated in the city of Graz, Austria in a relatively dense urban area. The school provides education for approximately 300 pupils between the ages of 6 and 10 years. The school was built in 2019 (nearly zero energy building standard) and is equipped with a $CO_2$-controlled ventilation system, concrete core activation for heating and cooling and automated shading devices. Opening the windows is possible. A centralized BMS was installed and all rooms are equipped with customary room control units and measurement devices (Siemens QMX3.P70). Intentionally, no additional measurement equipment was used for carrying out this pre-study. The heating system (district heating, heat pumps) was already activated during the feedback period (12.09.2022-28.10.2022).

As we present the results from an Austrian school, we have to consider Austrian laws and regulations. Table 1 provides a non-exhaustive overview of the applicable common regulatory framework and guidelines. It shows a variety of applicable indicators (partially voluntary) and some inconsistencies. Consider, for instance, operative temperature vs. air temperature, and average vs. median, among the criteria. Finally, note that as long as the minimum requirements are met (in case of legal disputes), the stakeholders, such as the building owner, planner, or architect, may agree upon differing design criteria. We want to highlight that although CO2 concentration levels are commonly used as the main IAQ parameter, several other pollutants (created indoors or outdoors) influence occupants’ perception of IAQ. The responsible German and Austrian ministries [34, 35] published comprehensive guidelines, including benchmarks, measurements, analysis, and detailed descriptions of single pollutants such as volatile organic compounds, formaldehyde, or phenol. Investigating these parameters, however, is beyond the scope of this paper.

Table 1. extract of relevant Austrian legislation and guidelines

| source | Name | Application | Indicators |
|---|---|---|---|
| [30] | ÖNORM B 8110-6-1:2019 | Calculation of heating energy demand and cooling energy demand for energy performance certificate of a building (legally binding based on Federal regulations) | indoor temperatures (not perceived temperature): minimum 22°C in heating season, maximum 26°C in cooling season |
| [31] | ÖNORM H 12831-1 | Heating systems in buildings - Method for calculating the standard heating load | indoor temperature 18-22 °C in schools, depending of type of room (e.g. 22°C in classrooms); valid if air temperature and operative temperature are identical |
| [32] | Austrian Workplace regulation; ensuring decent working conditions and workplaces for employees | §28 required air temperatures | air temperature 19-25°C in case of low physical activity or 18-24°C in case of moderate physical activity; if air handling unit is installed: < 25°C during cooling period |
| | | §27 humidity | In case of air handling unit: 40-70% |
| | | §27 ventilation | in case of installed ventilation system: outdoor air supply 35m³/h per adult (low physical activity) respectively 50 m³/h (moderate physical activity) |
| | | §25 daylight ingress | sum of light entry areas at least 10% of the floor area of the room |
| | | §29 artificial lightning | Minimum 100 lux (measured 0,85m above ground) |
| [16] | Position paper on ventilation of school and classroom spaces - SARS-CoV-2 | Official recommendations for ventilation of school and classroom spaces of the Federal Austrian ministry | outdoor air supply 35m³/h per adult; <1000ppm $CO_2$ (mean value of 1-minute instantaneous values during the assessment period) |
| [11] | official IAQ guideline provided by the Austrian Ministry | Guideline for the evaluation of indoor air quality - Carbon dioxide as a ventilation parameter. | <1000ppm $CO_2$ (mean value of 1-minute instantaneous values during the assessment period) |
| [33] | Klimaaktiv building certification for energy efficient buildings | Voluntary Austrian building certification system, applied on the case study | Median $CO_2$ levels below 1000ppm |

## Data collection

*Operational Data*

The building operator provided time series data from the building HVAC system through data exports in the .csv format. Operational data includes operative room temperatures (°C) and CO2 levels (ppm) for each room (15-min resolution on average). As per the configuration of the BMS, there are no humidity measurements. The building houses 29 rooms, 14 of which we included in the study. The 14 rooms are located across three floors, with three rooms on the first, nine on the second, and two on the third floor.

*Daily/Hourly User Feedback*

For the comfort study, we used tablets running a custom-built mobile application to collect feedback from the teachers. After an introductory meeting where we presented the app and the questionnaires, we asked the teachers to provide hourly and daily feedback on school days. Daily feedback includes information about odors and drafts, the quality of natural and artificial lighting, environmental noises as well as the teachers' perceptions of the pupils' levels of agitation, concentration, motivation, and how easy it was for the pupils to adhere to the rules and requirements of the lessons. Hourly feedback asks the teachers for an assessment of the overall room climate, temperature, humidity, and air quality. You can see a screenshot of the app's user interface (UI) in Figure 1 (daily feedback survey).

On an hourly basis, the following questions, using a 5-part scale were asked:

- How do you perceive the indoor climate in this room? (1 = very uncomfortable, 5 = very comfortable)
- How do you perceive the room temperature? (1 = very cold, 5 = very warm)
- How do you perceive humidity in this room? (1 = very dry, 5 = very humid)
- How do you perceive the air quality in this room? (1 = very stuffy, 5 = very sticky)

As already mentioned, the daily feedback survey was complemented with questions related to the perceived pupils' performance, again using a 5-part scale:

- How did you perceive the odor in the room today? (1 = very unpleasant, 5 = very pleasant)
- How did you perceive the air draft in the room today? (1=very drafty, 5 = no draft
- How did you perceive the sunlight in the room today? (1=very unpleasant, 5 = very pleasant)
- How did you perceive the artificial light in the room today? (1=very unpleasant, 5 = very pleasant)
- How did you perceive the environmental noise in the room today? (1 very loud, 5= very quiet)
- The pupils were agitated today: 1= very agitated, 5 = not at all
- The pupils were concentrated today: 1= not at all, 5 = very concentrated
- The pupils were motivated today: 1= not at all, 5 = very motivated
- It was easy for the pupils to adjust their behavior to rules: 1= very difficult, 5 = very easy

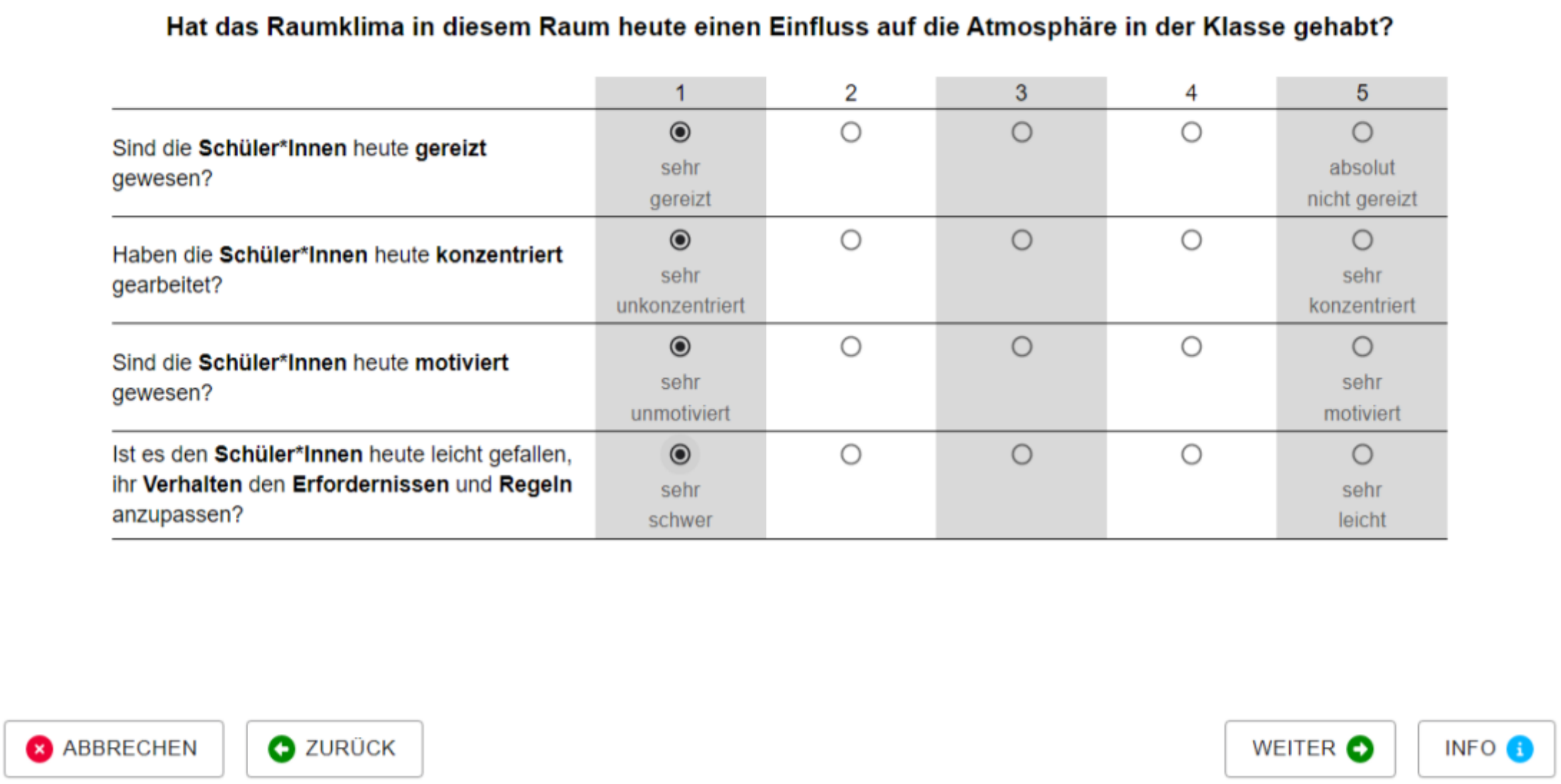

Figure 1. screenshot of the UI components for teachers' daily feedback related to the perceived performance of the pupils

The application is built around a client-server setup providing a database with an administration backend and a web frontend. The code for the application is available on request. A conceptual overview of the server/client architecture can be seen in Figure 2.

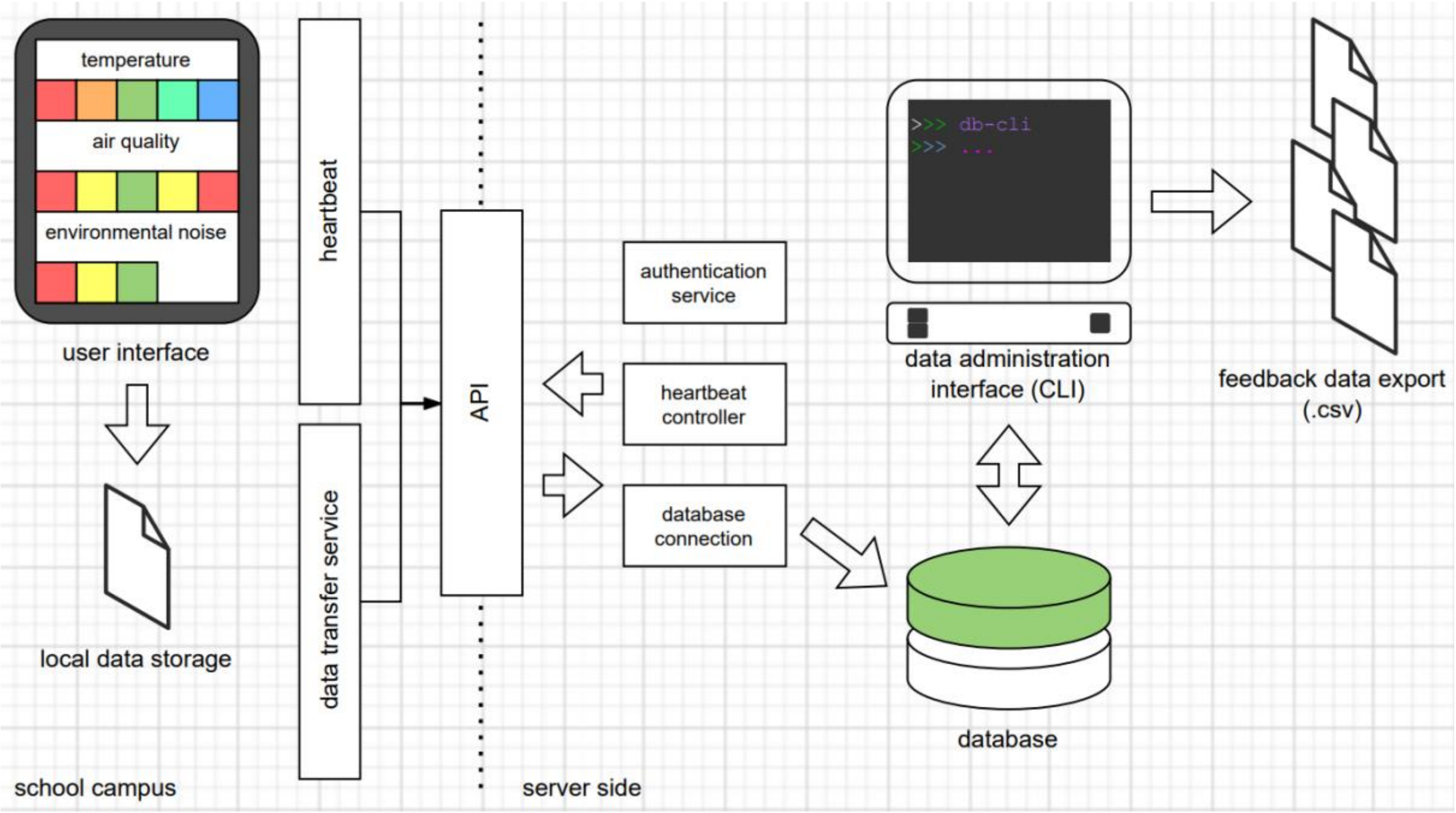

Figure 2. conceptual overview of the server/client architecture

We can categorize the application's components into client-side and server-side components, with an API connecting them. The client side consists of the UI, a client backend handling local data backup, data transmission, and a heartbeat service. The UI is a simple web-page-style application where occupants can enter their feedback. The client backend buffers feedback data locally until it can send it to the server. Local backup ensures that no data is lost in case of a network outage. The heartbeat service allows us to monitor the state of the device and detect problems, such as network or battery issues. The server side of the pipeline provides the functionality for the endpoints in the API, i.e., authentication, receiving, and processing of feedback data and heartbeat updates. The server hosts a document-based mongoDB[1] as well as data handling utilities. These utilities include a command line interface (CLI) for performing CRUD operations on the data and for exporting the data. To deploy the client bundle on the tablets, we used the Fully Kiosk app[2], Android Kiosk browser, and app launcher. Fully Kiosk provides a variety of functionalities, including a full-screen web browser that displays the client UI and prevents users from accessing other functionalities on the tablet. Additionally, we used Fully Kiosk's cloud functionality to remotely provision, monitor and configure the devices.

We set up the feedback collection pipeline with modular components, which means that the server side can easily be replaced with another backend, such as an IoT middleware platform, by simply having it implement the API. This architecture allows for seamless integration with building operation systems and ontologies as well as high-level building services, such as HVAC controllers or predictive control.

### Data analysis and correlations

In total, we collected 470 hourly feedback entries and 163 daily feedback entries from 22 teachers. For the daily feedback, we were interested in the overall trends for the entire day, explicitly considering how air quality and temperature and subsequently students' performance develop throughout the day. However, it is not uncommon for teachers to only be in school to supervise arriving children before classes start on some days hence we decided to exclude any daily feedback submitted before 8 AM. We did not explicitly consider gender-related differences since we only received five feedback entries from male teachers.

For the data analysis, we used Python-based notebooks, implemented using the open-source scientific computing libraries *NumPy*[3] and *pandas*[4] as well as the data visualization libraries *Matplotlib*[5] and *seaborn*[6]. The data pre-processing performed in this project can be split into four steps:

1. Loading the datasets into *pandas DataFrames*
2. Aligning the different datasets according to spatial and temporal criteria
3. Analyzing the sensor and feedback data including plausibility and consistency checks
4. Extracting statistics

We correlate the feedback data with sensor readings from the room where we recorded the feedback. For daily feedback entries, we use the mean over sensor readings from the beginning of the school day (8 AM) up until the timestamp of the feedback entry. For the hourly feedback, we use the mean over readings from the hour preceding the feedback submission.

For the correlation analysis, we used the Spearman coefficient, which quantifies the monotonic relationship between two variables that do not assume a normal distribution [55]. The p-value is a measure of probability that supports the correlation coefficient when determining the strength of the relationship between two variables. We determined the relevant significance level at <0,05. [56]

[1] https://www.mongodb.com/

[2] https://www.fully-kiosk.com/

[3] https://numpy.org

[4] https://pandas.pydata.org

[5] https://matplotlib.org

[6] https://seaborn.pydata.org

## RESULTS AND DISCUSSION

### Evaluation of indoor climate (operative temperatures, $CO_2$ levels)

The results for school days in the feedback period (12.09.2022-28.10.2022) show that during school hours (08:00-17:00 CEST), the average CO2 levels were between 508ppm and 975ppm in the 14 rooms (25% quartiles 439-797 ppm, 50% quartiles 487-874 ppm, 75% quartiles 526-1142ppm). Additionally, the measurements revealed higher CO2 concentrations in 8 classrooms (median 686-874ppm) compared to the other rooms (leisure rooms, music room, group rooms). For detailed results, see Table 3. The average indoor temperature varied between 23,3°C and 25.1°C in the 14 rooms (25% quartiles 23.0-24.4°C, 50% quartiles 23.3-24.8°C, 75% quartiles 23.9-25.3°C) during school hours. Table 2 provides a detailed analysis of the indoor temperature.

Table 2. overview on data analysis results for operative temperatures in the classrooms, 12.09.2022-28.10.2022

| Floor | Room | Data points | Mean operative temperature (°C) | Standard deviation (°C) | 25 % quartiles (°C) | 50 % quartiles (°C) | 75 % quartiles (°C) |
|---|---|---|---|---|---|---|---|
| 1 | leisure room | 1694 | 23,9 | 0,9 | 23,3 | 23,9 | 24,4 |
| 1 | music room | 1701 | 23,3 | 0,9 | 22,7 | 23,3 | 23,9 |
| 1 | leisure room | 1700 | 23,9 | 1,2 | 23,1 | 23,8 | 24,6 |
| 2 | classroom | 1696 | 24,1 | 0,8 | 23,5 | 24,0 | 24,6 |
| 2 | classroom | 1700 | 24,3 | 0,7 | 23,9 | 24,4 | 24,8 |
| 2 | classroom | 1700 | 24,0 | 0,8 | 23,4 | 24,0 | 24,5 |
| 2 | group room | 1700 | 24,8 | 1,3 | 23,9 | 24,6 | 25,3 |
| 2 | classroom | 1699 | 24,3 | 1,0 | 23,6 | 24,3 | 25,0 |
| 2 | classroom | 1700 | 23,6 | 1,0 | 23,0 | 23,8 | 24,4 |
| 2 | classroom | 1698 | 23,8 | 0,9 | 23,0 | 23,8 | 24,4 |
| 2 | classroom | 1698 | 23,4 | 0,7 | 23,0 | 23,4 | 23,9 |
| 2 | group room | 1698 | 24,1 | 1,5 | 23,2 | 23,8 | 24,5 |
| 3 | classroom | 1699 | 24,2 | 0,9 | 23,5 | 24,3 | 24,9 |
| 3 | group room | 1699 | 25,1 | 1,1 | 24,4 | 24,8 | 25,5 |

Table 3. overview on data analysis results for $CO_2$ levels in the classrooms, 12.09.2022-28.10.2022

| Floor | Room | Data points | Mean $CO_2$ concentration (ppm) | Standard deviation (ppm) | 25 % quartiles (ppm) | 50 % quartiles (ppm) | 75 % quartiles (ppm) |
|---|---|---|---|---|---|---|---|
| 1 | leisure room | 1696 | 508 | 110 | 439 | 487 | 526 |
| 1 | music room | 1701 | 588 | 177 | 451 | 528 | 684 |
| 1 | leisure room | 1700 | 535 | 112 | 460 | 501 | 583 |
| 2 | classroom | 1699 | 975 | 245 | 797 | 874 | 1142 |
| 2 | classroom | 1701 | 950 | 218 | 793 | 857 | 1104 |
| 2 | classroom | 1699 | 765 | 302 | 498 | 704 | 957 |
| 2 | group room | 1700 | 880 | 165 | 783 | 825 | 914 |
| 2 | classroom | 1698 | 819 | 288 | 539 | 786 | 1022 |
| 2 | classroom | 1699 | 771 | 243 | 539 | 746 | 945 |
| 2 | classroom | 1698 | 849 | 334 | 533 | 776 | 1069 |
| 2 | classroom | 1698 | 815 | 287 | 536 | 768 | 1021 |
| 2 | group room | 1697 | 792 | 265 | 532 | 759 | 974 |
| 3 | classroom | 1699 | 748 | 280 | 476 | 686 | 964 |
| 3 | group room | 1699 | 619 | 181 | 478 | 579 | 739 |

As outlined in Table 1, the common design criterion for indoor temperature is a minimum of 22°C for the heating period, whereas during the operation phase, the Austrian Workplace Regulatory §28 considers an air temperature range between 19-25°C (rooms with low physical activity) respectively 18-24°C (rooms with normal physical activity) throughout the year as adequate. Table 2 shows that the averaged measured room temperatures (23.1-25.1°C) are on the upper limit of the recommended ranges. It needs to be mentioned that the shown temperatures in Table 2 are the measured operative temperatures. There were no separate air temperature measurements available. However, the Austrian Workplace Investigation Unit [57] assumes that if operative temperatures comply with the provided air temperature ranges defined in the Workplace Regulatory, it is very likely that the indoor climate is acceptable for employees. However, our analysis of the temperature ranges is in line with the feedback from the teachers and shows that room temperatures are on the higher end of the spectrum (see section “hourly feedback data evaluation”). In our study, we did not collect feedback from the pupils. However, it is likely that the indoor temperatures during the survey period are not adequate for the pupils either. As outlined in the background section, evidence suggests that children have different thermal perceptions than adults, resulting in lower comfort temperatures.

In addition to thermal comfort issues, recent investigations such as [19] show a relationship between classroom temperatures and performance, predicting that reducing temperature from 30°C to 20°C is most likely to increase the performance of school work by 20%. Opposed to the ideal temperature for effective performance in offices, the indoor temperature in schools should be lower. In line with the studies cited in the section “Introduction and related background”, the explanations above support the theory that the design of educational buildings and their indoor environment have to change.

Besides operative temperatures, we analyzed CO2 levels. Figure 2 displays the daily course of the CO2 concentrations in all classrooms for the feedback period. Assuming that classes start at approximately 08:00 CEST, CO2 levels rise. This trend confirms that occupants are the main source of CO2. CO2 concentration peaks after 2-4 hours of teaching and decreases during the afternoon. Since our investigation took place in a primary school and the teachers provided approx. 60% of the daily feedback entries before 12:00 CEST, it is likely that the decreasing CO2 concentrations are mainly caused by the absence of occupants and not by adapted control measures of the centralized ventilation system.

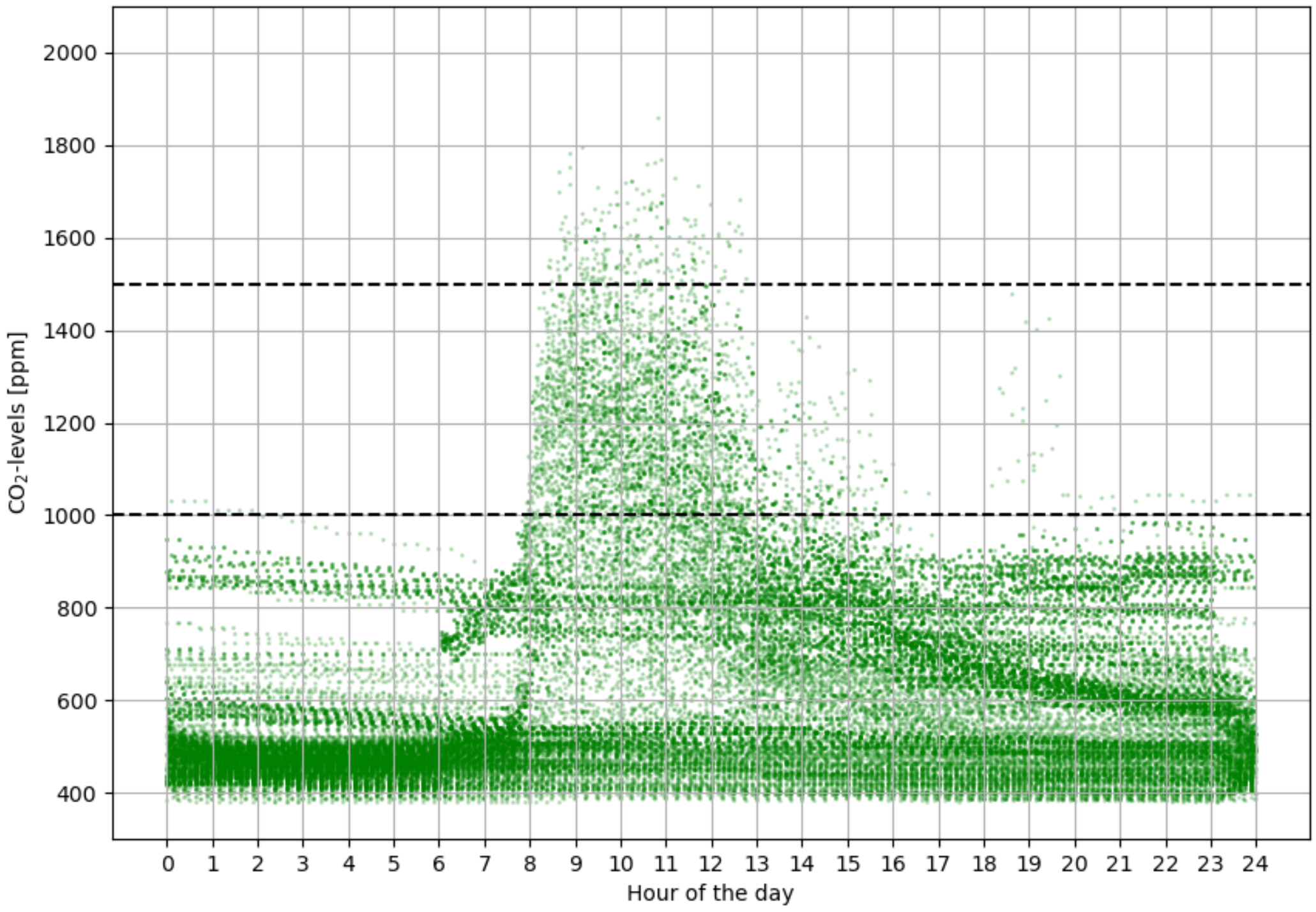

Figure 2. daily course of the $CO_2$ concentrations in the 14 rooms for the feedback period (12.09.2022-28.10.2022)

Approximately 20% of all measurement values during school hours were above 1000 ppm, and 2% were above 1500 ppm. Since CO2 levels can change quite quickly (e.g. opening of windows), these rates might be even higher if data was of higher resolution (e.g. 1-minute values as recommended by relevant guidelines [16]). According to Umweltbundesamt (n.d.), CO2 concentrations above 1000 ppm are hygienically critical, which is especially relevant after the recent experiences with the COVID-19 pandemic [34]. Centralized HVAC systems must be designed to keep average CO2 levels below 1000 ppm during the lectures [35]. However, some researchers [58] and institutions [59, 60], such as the official U.S. Center for Disease Control and Prevention propose a CO2 threshold of 800 ppm to reduce the risk of airborne disease infections. According to McLeod et al. (2022), it might be challenging to achieve such thresholds in existing buildings, because of a threefold increase in required ventilation rates for typical classrooms and the risk of causing lower thermal comfort throughout the year, and increased energy demand in winter times [15].

Based on a six-day physical environment measurement and learning performance test at Zhejiang Sci-Tech University, Ma et al. (2023) concluded that indoor air CO2 concentrations below 700 ppm significantly improve all learning abilities, including memory and logic deduction [61]. Bakó-Biró et al. (2012) found that CO2 concentrations below 1000 ppm ensure an effective learning environment [20]. Our analysis confirmed the need to adapt design requirements for school buildings so that they acknowledge the importance of high IAQ for providing an optimum learning environment.

**Hourly feedback data evaluation**

Figure 3 and Figure 4 show the scatter plots, Spearman correlations, and p-value between the average hourly measurement data for CO2 levels and operative temperatures and the hourly feedback from the teachers in the 14 rooms. We used approximately 330-430 data points to analyze the variables. Plot (a) in Figure 3 shows that the majority of the comfort votes were on the upper range of the scale (1 = very cold, 5 = very warm). Approx. 44% of all feedback votes in the 14 rooms were either “4” or “5”.

The Spearman correlation coefficient of 0.23 (p-value <0,05) suggests a weak positive correlation between the operative room temperature and the perceived indoor temperature of the occupants (a). We can conclude that occupants tend to vote that it is too warm if the operative room temperature increases. The apparent preferred temperature range (feedback vote “3”) is quite broad (19.3-25.6°C). It demonstrates the variety of individual comfort preferences. The room temperature seems to influence the perceived overall IAQ (weak negative correlation of -0,13) (Figure 3, b), indicating that lower indoor temperatures are associated with better IAQ.

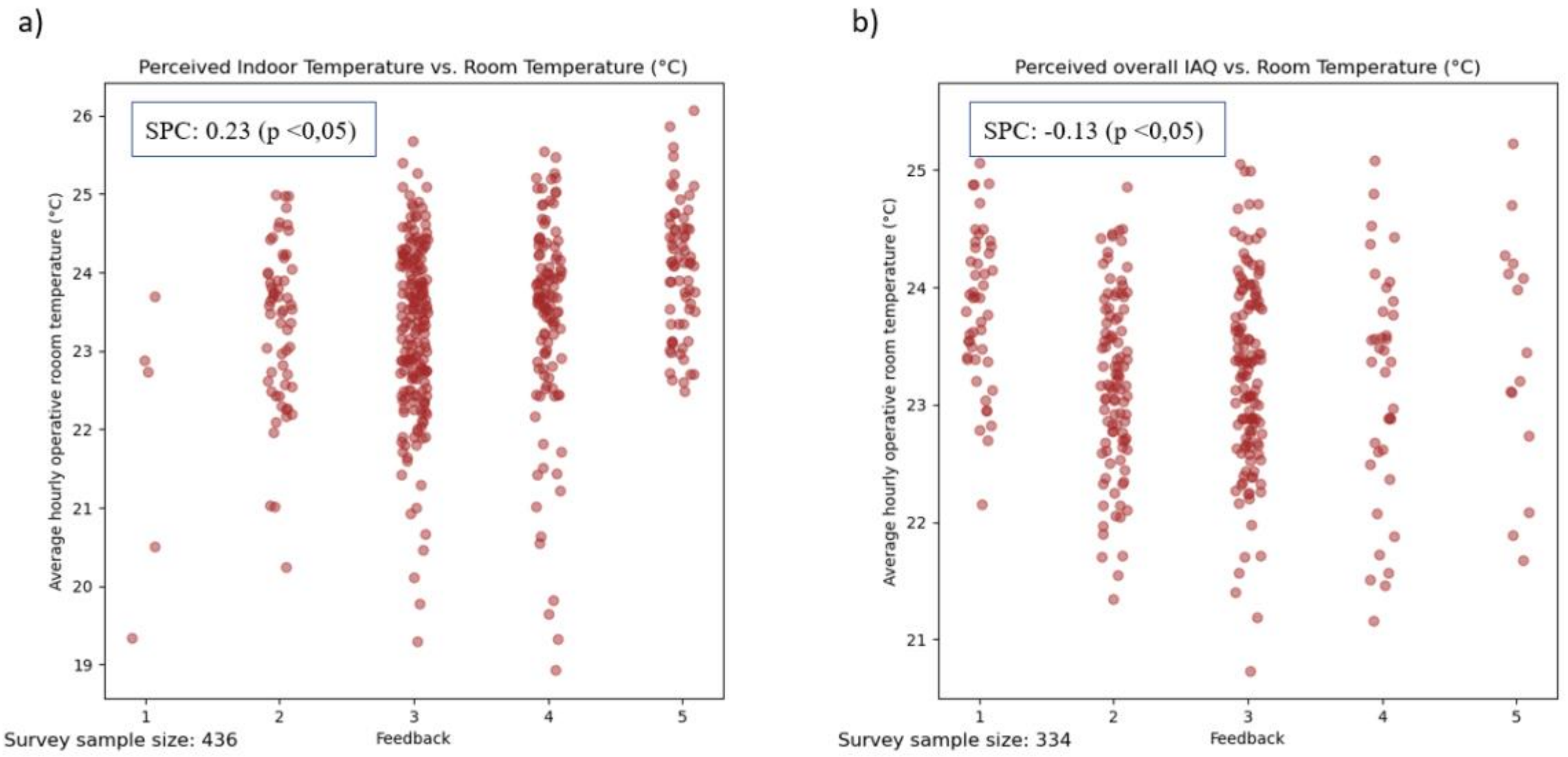

Figure 3. Results of correlation analysis for teachers’ feedback vs. room temperature: x-axis feedback scale in (a) 1 = very cold, 5 = very warm; x-axis feedback scale in (b) 1 = very uncomfortable, 5 = very comfortable

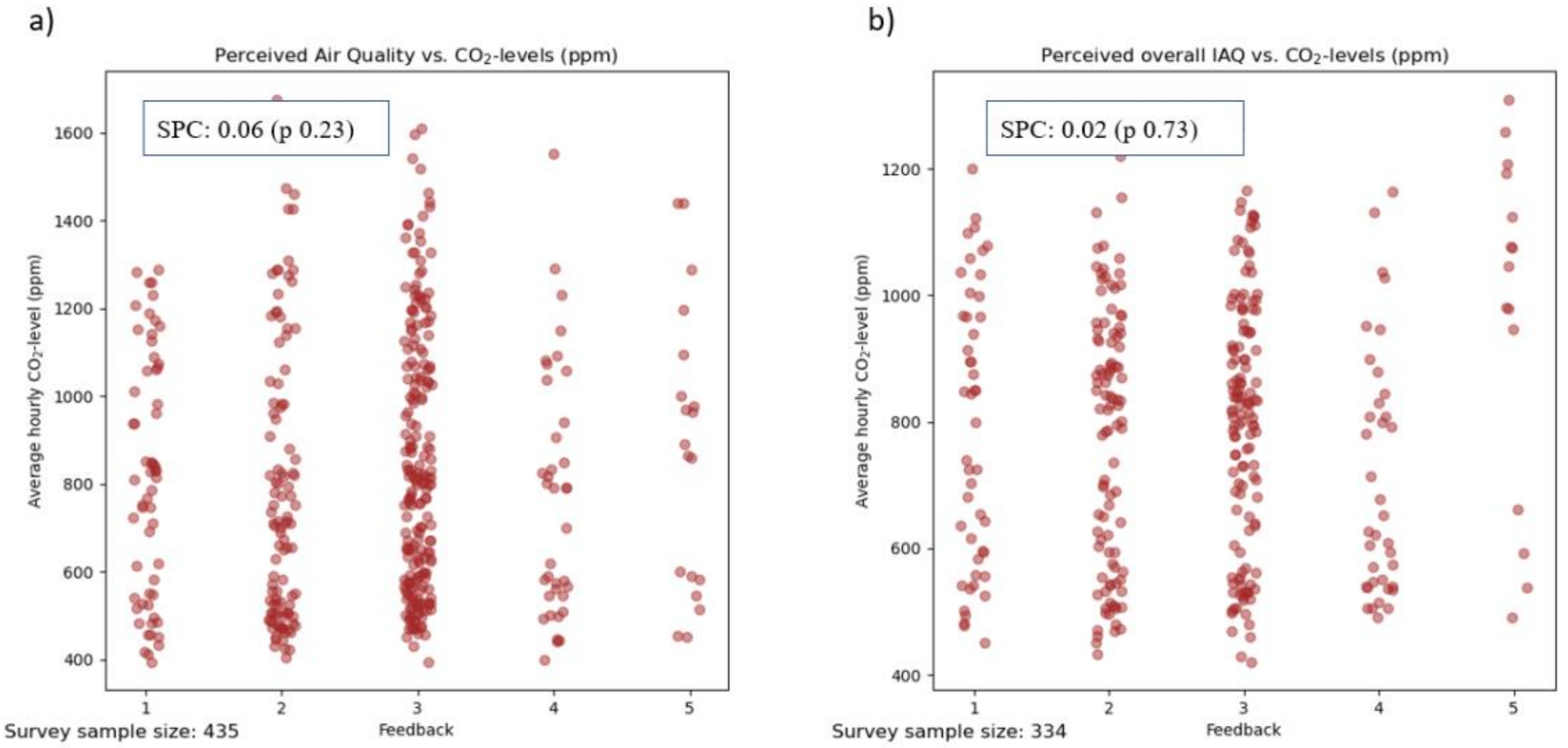

Figure 4. Results of correlation analysis for feedback vs. CO2 levels: x-axis feedback scale in (a) 1 = very stuffy, 5 = very sticky; (b) 1 = very uncomfortable and 5 = very comfortable

Note that drops in indoor temperatures could have been caused by open windows, which influences other IAQ parameters such as humidity levels and CO2 levels in the rooms. However, the BMS did not collect humidity measurements, and there was no statistically significant correlation between perceived IAQ and CO2 levels (Figure 4, b). Additionally, we did not find any significant correlations between perceived air quality and CO2 levels (Figure 4, a). There are several reasons why there were no strong correlations between comfort feedback and temperature and CO2 levels. Some possible reasons include:

- Complexity of comfort: as already outlined in the background section, comfort is a multidimensional concept that can be affected by a variety of factors including temperature, humidity, air quality, noise, lighting, and more. The lack of correlation between comfort feedback and temperature and $CO_2$ levels supports this theory.
- Adaptation: Zhang et al. (2023) showed that people tend to adapt to their environment over time, which can affect their perception of comfort [50]. The teachers may have become accustomed to the indoor conditions in the school, making it difficult to detect a correlation between the comfort feedback and the measurement data.
- Variability in the data: The comfort ratings provided by the teachers may have been affected by the day-to-day variability of the teaching schedule, class size, and other factors, which can make it difficult to detect a correlation with the measured temperature levels and $CO_2$ levels.
- Other factors affecting comfort: There may be other factors such as the design of the building, the type of furniture, the presence of natural light, or the quality of the acoustics that are impacting the teachers' perceptions of comfort and that are not captured in the temperature and $CO_2$ measurements.
- The design of the study or measurement errors might contribute to obtaining weak correlations.

**Daily feedback data evaluation**

As advised by Lamberti et al. (2021), IEQ aspects were included in the feedback survey [62]. As already described in the method section, we asked 22 teachers to provide daily feedback after their last lesson related to room climate (perceived odor, draft, noise, daylight, and artificial light) during this day using a 5-point scale. We correlated feedback entries and the average temperature and CO2 measurement data for this day (08:00 CEST until the timestamp of feedback entry). The teachers provided approximately 60% of the daily feedback entries before 12:00 CEST. This trend is plausible in a primary school setting. Additionally, changing the room for the next lesson might be a reason for early feedback entries.

Figure 5 shows the result for the calculated Spearman correlation (SPC) and p-value for odor and measurement data (156 available data points). Whereas perceived odor and CO2 measurement data show no statistically significant correlation (p-value 0.1; SPC -0,13), the room temperature and perceived odor have a moderate negative correlation of -0,31. This indicates that lower room temperatures result in a more pleasant odor perception.

Zhang et al. (2023) reported similar findings. Perceived odor intensity in classrooms was significantly lower at 24 °C compared to 27°C indoor temperature, accompanied by higher satisfaction with indoor air quality. The air quality inside the classroom was deemed acceptable for occupants at both 24 °C and 27 °C if the CO2 concentration remained below 2.600 ppm [50].

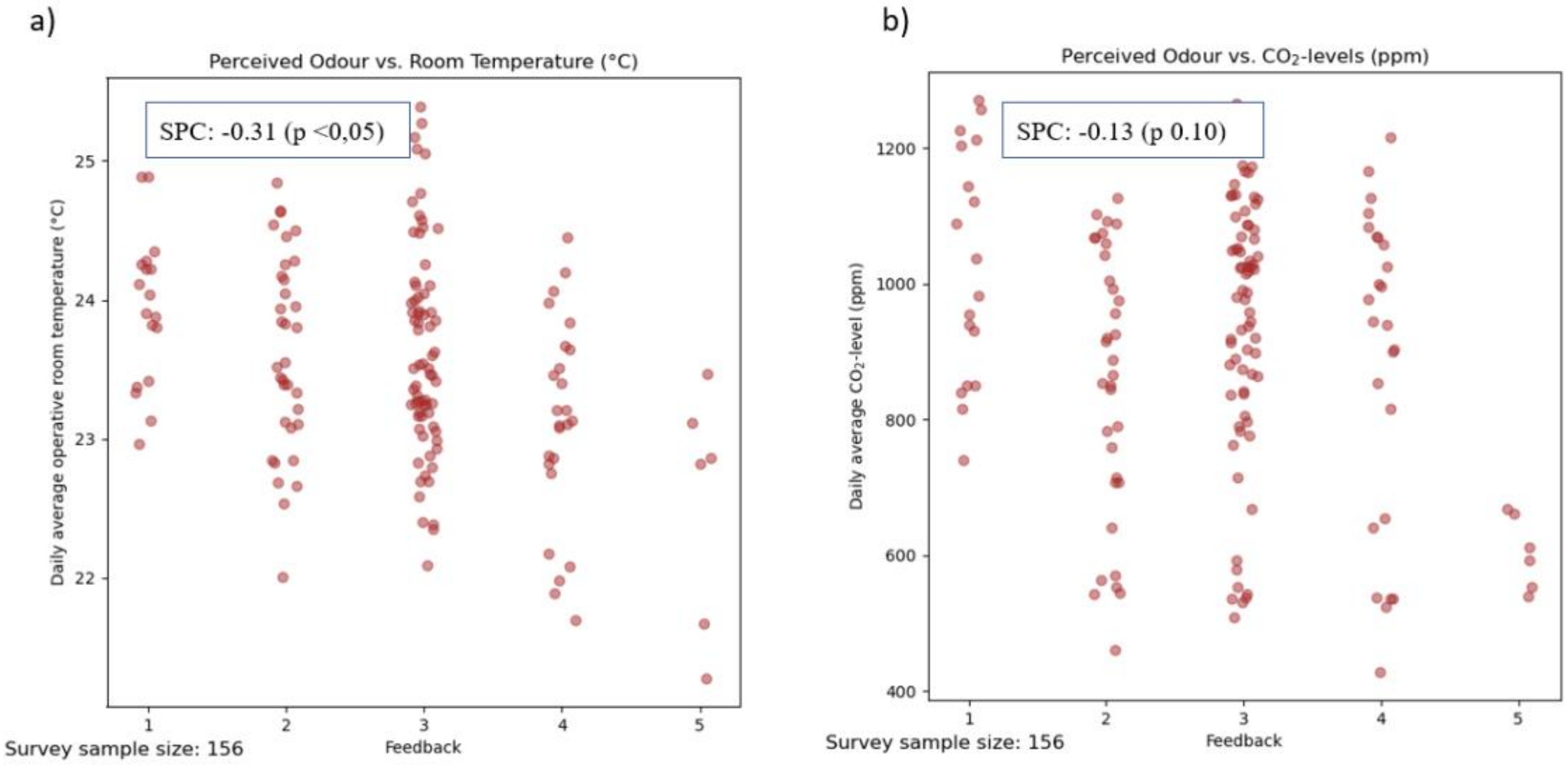

Figure 5. Results of correlation analysis for perceived odor vs. measurement data: x-axis feedback scale in (a) & (b) 1 = very unpleasant, 5 = very pleasant

Besides IEQ-related parameters, teachers were asked to evaluate the performance of the pupils, using subjective indicators such as motivation, focus, adherence, and irritation. No or very weak correlations (<0,8; without statistical significance) were found between the teachers' performance ratings and the daily average indoor temperature and CO2 levels. Additionally, we performed an analysis of all of the teachers' feedback entries. The results of the Spearman correlation analysis are shown in Figure 6 below. Moderate significant correlations (0.22 – 0.39) were found between IEQ indicators and performance indicators, highlighting that "ambient sounds" and lightning situations are perceived to have a relevant impact on pupils' behavior. Similar findings are shown in the literature.

Mudenda et al. (2021) indicated that inadequate illumination causes discomfort and adversely affects the academic performance of undergraduate pharmacy students. The quality of lighting in classrooms could potentially influence the grades [63]. Hyiid et al. (2020) investigated this hypothesis to examine the impact of lighting and ventilation in students' homes on their academic performance, using five statistical scenarios. The study involved ninety-two children aged between 10 to 12 years old, and data was collected over a period of four weeks, evaluating children's cognitive abilities, such as processing speed, focus, reasoning, and problem-solving skills. Results show better performances in terms of processing speed, concentration, and math skills. However, the authors admit that there was not a statistical significance related to the exact contribution of the individual measures (improved lightning or high ventilation rate) but the d2-test results indicate that both measures applied individually can have the same effect on the pupil's cognitive performance[64]. Besides these uncertainties, further empirical investigations are necessary related to the causes (indoor, and outdoor) of noise pollution in schools located in urban areas and their effects on children's behavior.

IEQ parameter "odor" correlates with perceived "focus" and "irritation" (0.33 and 0.31). Since perceived "odor" is influenced by factors such as room temperature [50], evaporations of humans, inventory, building materials, mold, etc. in the classroom [65], perceived "odor" can be a result of a lack of fresh air and therefore of low ventilation rates. As shown before, Figure 5 b) indicates a weak negative correlation between perceived odor and average daily CO2 measurements. However, since this correlation is not significant (p-value 0,1), further investigations about the odor sources are necessary to exclude potentially harmful substances (evaporating materials, mold, etc.). Figure 6 shows no influence of “air draft” on perceived pupils’ performance, which most likely results from the absence of air draft in this school building, built in 2019 (high-quality building envelope).

The strong significant correlations >0.66 between the performance ratings themselves indicate consistency within the feedback entries. Perceived focus and adherence as well as motivation seem to be closely related, which might result from natural behavioral similarities among these characteristics. These subjective ratings of the teachers were not compared against measurable performance indicators such as school tasks or psychological tests [20,66], which might be a future research direction to compare different learning environments (e.g. nearly zero energy buildings and older buildings). In addition, future studies need to concentrate on the therapeutic potential of indoor environments that encompass all aspects of indoor environmental quality (IEQ), including thermal comfort, air quality, visual quality, and acoustic quality. As individuals are exposed to a combination of these factors, analyzing their combined impact is essential to comprehending human perception [62].

| | *chosen IEQ indicators* | | | | | *performance indicators* | | | |
|---|---|---|---|---|---|---|---|---|---|
| | air draft | ambient sounds | artificial lights | daylight | odour | adherence | focus | irritation | motivation |
| adherence | 0,04 | 0,38 | 0,25 | 0,35 | 0,26 | 1,00 | 0,79 | 0,69 | 0,66 |
| focus | 0,03 | 0,35 | 0,35 | 0,34 | 0,33 | 0,79 | 1,00 | 0,76 | 0,77 |
| irritation | 0,05 | 0,22 | 0,30 | 0,26 | 0,31 | 0,69 | 0,76 | 1,00 | 0,72 |
| motivation | 0,00 | 0,24 | 0,39 | 0,30 | 0,25 | 0,66 | 0,77 | 0,72 | 1,00 |

Figure 6. Spearman correlations for IEQ ratings and performance ratings of the teachers (160 feedbacks); except air draft indicator, all correlations are statistically significant (p-values <0,05)

## LIMITATIONS OF THE STUDY

With this pre-study we are able to provide a data set on indoor air quality including both numeric sensor data and empirical occupant assessments. It is important to note that the study was designed to give an abstract, high-level indication of IEQ and IAQ throughout the entire school building, aggregated over the whole feedback period, rather than detailed in-depth-measurements of chosen rooms at chosen times. Hence, we cannot infer strong correlations between pupils’ performance, perceived and measured room climate without further investigation.

As per the study’s design we did not collect influence factors, such as clothing or activity rate that are typically used in conventional comfort models such as PMV and instead focused on correlating the feedback with measurements from the BMS directly. While humidity would have been an important factor in the assessment of IEQ and IAQ, unfortunately the school’s BMS did not provide such measurements, as they are not part of the HVAC’s control policy. Without direct access to the BMS we were also limited not only in terms of the quantities of interest but also in terms of the data’s resolution.

Finally, we want to highlight that we deliberately chose to evaluate the influence of the IEQ and IAQ on the pupils indirectly, i.e., through the teachers’ assessment of their performance instead of their own perception.

## CONCLUSION AND LESSONS LEARNED

In this paper, we implemented and tested an occupant feedback data collection framework and investigated correlations between user comfort and operational characteristics of a school building over the course of two months in September and October 2022. To collect hourly and daily assessments of the teachers' comfort, we installed tablets running a simple web-based feedback application in several rooms of the building. We collected feedback data in a database and aligned it with the temperature and $CO_2$ data we obtained from the BMS.

Initial results show that during the feedback period (12.09.2022-28.10.2022) the average room temperatures (23.1-25.1°C) in all rooms are at the upper limit of the recommended temperature ranges, which is in line with the feedback from the teachers. Our literature review demonstrated that comfort temperatures for children are lower than for adults. Hence, it stands to reason, that the indoor temperatures during the survey period are not adequate for the pupils either. Our analysis showed that during school hours, approximately 20% of all $CO_2$ measurement values were above 1000 ppm, and 2% were above 1500 ppm. $CO_2$ levels above 1000 ppm are considered to be hygienically critical and ventilation systems should be able to adapt the ventilation rates to prevent such high concentrations. However, the latest research proposes lowering the limits below 800 ppm to ensure a healthy and effective learning environment.

We did not find strong correlations between the hourly feedback and the measurement data. While it is difficult to pinpoint why, possible explanations include the complex nature of comfort or people's adaptation capabilities. We also collected daily feedback on IEQ (perceived odor, draft, noise, daylight, and artificial light). We performed a correlation analysis between these IEQ indicators and average indoor temperatures and $CO_2$ levels. Whereas perceived odor and $CO_2$ measurement data show no statistically significant correlation, the room temperature and perceived odor have a moderate negative correlation of -0,31 (p-value $<0,05$). This indicates that lower room temperatures result in a more pleasant odor perception. We propose to investigate the sources of the odor to exclude potentially harmful substances (evaporating materials, mold, particulate matter, volatile organic compounds, etc.). Additionally, we asked the teachers to rate the performance of the pupils. We found no strong correlations between perceived performance and measurement data. Strongly significant correlations $>0.66$ within the performance ratings indicate consistency. Perceived focus and adherence as well as motivation seem to be closely related, which results from natural behavioral similarities among these characteristics. We did not complement the subjective ratings of the teachers with performance tests. This might be a future research direction to compare different learning environments (e.g. nearly zero energy buildings and older buildings).

Overall, we conclude that further investigation would be needed to determine the exact cause of the correlations and non-correlations and to identify potential solutions for improving the comfort of the occupants. Additionally, it would be interesting to collect feedback from pupils to retrieve their comfort perceptions and activity levels and combine them with adequate, measurable performance tests. Besides this, repeating the study in the middle of the heating period (e.g. December) or cooling season (e.g. June) might encourage even more beneficial insights.

The use of separate, modular components for the feedback collection pipeline provides us with a flexible framework for occupant participation; and while our setup fulfilled the requirements of this pre-study, there is potential for improvement. Connecting the feedback components into existing building automation systems, for instance, would allow operators to directly align feedback data with operational data in the automation system, and to integrate signals from the occupants into the building control strategies. With the UI side of the feedback app being separated from the backend through the API such an integration would simply require an existing building automation system to provide the API endpoints to directly obtain the data from the tablets. Then there are many possibilities to process and utilize the data further, e.g., an intelligent HVAC control service could be programmed to take occupant comfort into account. Additionally, integrating the feedback components into building operation could potentially prevent issues during correlation analysis, as the system could be set to collect sensor readings whenever feedback is received, thus ensuring temporal alignment between the feedback and the measurements.

While we only assessed the challenges of providing a healthy indoor environment for an educational building in Austria, our literature review shows similar challenges and research efforts worldwide. In line with related research, our analysis demonstrates the need for adapting design requirements, especially for school buildings, acknowledging the different comfort needs of adults and children and the importance of high indoor air quality for providing an optimum learning environment. Future research should focus on testing adapted indoor environmental quality requirements for schools, especially in urban areas, and how to integrate real-time occupant feedback in the heating, ventilation and air conditioning systems.

## OPEN DATA

The data are freely available for research purposes at https://github.com/tug-cps/SchoolFeedbackStudyData.

## REFERENCES

1. European Commission (2019). Communication from the Commission to the European Parliament, the European Council, the Council, the European Economic and Social Committee and the Committee of the Regions – The European Green Deal. COM(2019) 640 final. Brussels
2. European Commission. (2020, February 17). Focus on energy efficiency in buildings. Retrieved from https://ec.europa.eu/commission/presscorner/detail/en/FS_20_218 [Accessed 15-03-2023]
3. European Parliament, & Council of the European Union. (2010). Directive 2010/31/EU of the European Parliament and of the Council of 19 May 2010 on the energy performance of buildings (recast). Official Journal of the European Union, L 153/13-L 153/35.
4. European Commission (2019). Proposal for a DIRECTIVE OF THE EUROPEAN PARLIAMENT AND OF THE COUNCIL on the energy performance of buildings (recast). COM(2021) 802 final. 2021/0426 (COD) Brussels
5. Pezzutto, S., Zambotti, S., Croce, S., Zambelli, P., Garegnani, G., Scaramuzzino, C., Pascuas, R. P., Zubaryeva, A., Haas, F., Exner, D., Müller, A., Hartner, M., Fleiter, T., Klingler, A.-L., Kühnbach, M., Manz, P., Marwitz, S., Rehfeldt, M., Steinbach, J., & Popovski, E. (2018). Hotmaps Project, D2.3 WP2 Report – Open Data Set for the EU28. Reviewed by Kranzl, L., & Fritz, S. (TUW). Hotmaps Project (Grant Agreement number 72367) https://gitlab.com/hotmaps/building-stock [Accessed 15-03-2023]
6. Knotzer, A., & Venus. D. (2013). School vent cool - Lüftung, passive Kühlung und Strategien für hochwertige Schulsanierungen. Bundesministerium für Verkehr, Innovation und Technologie. Vienna. Austria. p. 17
7. Becker, S. (2022). Metastudiezur Verbesserung der Datengrundlage im Gebäudebereich. Berlin. Munich. Germany. BMWK-Projekt-Nr.: 115/21-5, p. 9, p. 19-24
8. Jakob, M., Catenazzi, G., Melliger, M., Forster, M., Martius, G., Menard G. (2016). Potenzialabschätzung von Massnahmen im Bereich der Gebäudetechnik. Zürich. Switzerland p.17
9. Dorizas, P., De Groote, M. & Volt, J. (2018). Linking indoor environmental quality and energy performance in building regulation. Buildings Performance Institute Europe. Brussels. Belgium. p. 8 & p. 13
10. U.S. Environmental Protection Agency. (n.d.). Why Indoor Air Quality Is Important for Schools. https://www.epa.gov/iaq-schools/why-indoor-air-quality-important-schools [Accessed 15-03-2023]
11. Tappler, P. (2017). Richtlinie zur Bewertung der Innenraumluft – Kohlenstoffdioxid als Lüftungsparameter, Working Group on Indoor Air Quality at the Federal Ministry of Agriculture, Forestry, Environment and Water Management (BMLFUW). Vienna. Austria. p. 35
12. Awada, M., Becerik-Gerber, B., Hoque, S., O'Neill, Z., Pedrielli, G., Wen, J., & Wu, T. (2021). Ten questions concerning occupant health in buildings during normal operations and extreme events including the COVID-19 pandemic. Building and Environment, 188, 107480. https://doi.org/10.1016/j.buildenv.2020.107480
13. Kunkel, S., Kontonasiou, E., Arcipowska, A., Atanasiu, B., & Haines, V. (2015). Indoor air quality, thermal comfort and daylight: Analysis of residential buildings regulations in eight EU member states. Buildings Performance Institute Europe. Brussels. Belgium.

14. Crook, B., & Burton, N. C. (2010). Indoor moulds, Sick Building Syndrome and building related illness. Fungal Biology Reviews, 24(3-4), 106-113. https://doi.org/10.1016/j.fbr.2010.05.001
15. McLeod, R. S., Hopfe, C. J., Bodenschatz, E., Moriske, HJ., Pöschl, U., Salthammer, T., Curtius, J., Helleis, F., Niessner, J., Herr, C., Klimach, T., Seipp, M., Steffens, T., Witt, C., & Willich, S. N. (2022). A multi-layered strategy for COVID -19 infection prophylaxis in schools: A review of the evidence for masks, distancing, and ventilation. Indoor Air, 32(10), e13142. [e13142]. https://doi.org/10.1111/ina.13142
16. Tappler, P., Hutter, H. & Moshammer, H. (2020). Positionspapier zur Lüftung von Schul- und Unterrichtsräumen – SARS-CoV-2. Position paper of the Working Group on Indoor Air Quality. Federal Ministry for Climate Action, Environment, Energy, Mobility, Innovation and Technology (BMK). Vienna, Austria. P. 7, P. 37
17. Mendell, M. J., & Heath, G. A. (2005). Do indoor pollutants and thermal conditions in schools influence student performance? A critical review of the literature. Indoor Air, 15, 27-32.
18. Vornanen-Winqvist, C., Järvi, K., Andersson, M. A., Duchaine, C., Létourneau, V., Kedves, O., Kredics, L., Mikkola, R., Kurnitski, J., & Salonen, H. (2020). Exposure to indoor air contaminants in school buildings with and without reported indoor air quality problems. Environment International, 141, 105781. https://doi.org/10.1016/j.envint.2020.105781
19. Wargocki, P., Porras-Salazar, J. A., & Contreras-Espinoza, S. (2019). The relationship between classroom temperature and children's performance in school. Building and Environment, 157, 197-204. https://doi.org/10.1016/j.buildenv.2019.04.046
20. Bakó-Biró, Zs., Clements-Croome, D. J., Kochhar, N., Awbi, H. B., & Williams, M. J. (2012). Ventilation rates in schools and pupils' performance. Building and Environment, 48, 215-223. https://doi.org/10.1016/j.buildenv.2011.08.018
21. Chithra V.S. and Shiva Nagendra S.M. (2018). A Review of Scientific Evidence on Indoor Air of School Building: Pollutants, Sources, Health Effects and Management. Asian Journal of Atmospheric Environment, 12(2), 87-108. https://doi.org/10.5572/ajae.2018.12.2.87
22. Shendell, D. G., Prill, R., Fisk, W. J., Apte, M. G., Blake, D., & Faulkner, D. (2004). Associations between classroom CO2 concentrations and student attendance in Washington and Idaho. Lawrence Berkeley National Laboratory, Environmental Energy Technologies Division, Indoor Environment Department.
23. Julious, S.A., Osman, L.M., & Jiwa, M. (2007). Increases in asthma hospital admissions associated with the end of the summer vacation for school-age children with asthma in two cities from England and Scotland. Public Health, 121(6), 482-484. doi: 10.1016/j.puhe.2006.11.011.
24. Simoni, M., Annesi-Maesano, I., Sigsgaard, T., Norback, D., Wieslander, G., Nystad, W., Canciani, M., Sestini, P., & Viegi, G. (2010). School air quality related to dry cough, rhinitis and nasal patency in children. European Respiratory Journal, 35(4), 742-749. https://doi.org/10.1183/09031936.00016309
25. Tartarini, F., Schiavon, S., Quintana, M., & Miller, C. (2022). Personal comfort models based on a 6-month experiment using environmental parameters and data from wearables. Indoor Air, 32, e13160. https://doi.org/10.1111/ina.13160
26. Kim, J., Schiavon, S., & Brager, G. (2018). Personal comfort models – A new paradigm in thermal comfort for occupant-centric environmental control. Building and Environment, 132, 114-124. https://doi.org/10.1016/j.buildenv.2018.01.023

27. Xie, J., Li, H., Li, C., Zhang, J., & Luo, M. (2020). Review on occupant-centric thermal comfort sensing, predicting, and controlling. Energy and Buildings, 226, 110392. https://doi.org/10.1016/j.enbuild.2020.110392
28. de Dear, R.J., Akimoto, T., Arens, E.A., Brager, G., Candido, C., Cheong, K.W.D., Li, B., Nishihara, N., Sekhar, S.C., Tanabe, S., Toftum, J., Zhang, H. and Zhu, Y. (2013). Progress in thermal comfort research over the last twenty years. Indoor Air, 23, 442-461. https://doi.org/10.1111/ina.12046
29. Fabbri, M. & Dorizas, P. (2019): How to integrate indoor environmental quality within national long-term renovation strategies. Buildings Performance Institute Europe. Brussels. Belgium.
30. Austrian Standards Insitute (2019). ÖNORM B 8110-6-1:2019 - Building construction - Part 6-1: Indoor environmental quality - Indoor air quality - Requirements, recommendations and information. Austrian Standards International. Vienna, Austria.
31. Austrian Standards Institute. (2018). ÖNORM H 12831-1:2018-12-01. Daylight in buildings - Part 1: General aspects, terminology, principles and procedures. Austrian Standards International. Vienna, Austria.
32. Arbeitsstättenverordnung – AStV. Regulation of the Federal Minister for Labour, Health and Social Affairs, laying down requirements for workplaces and buildings on construction sites and amending the Construction Workers Protection Regulation. BGBl. II Nr. 368/1998 [CELEX-Nr.: 389L0654, 392L0057].
33. Federal Ministry for Sustainability and Tourism (2019). klimaaktiv Kriterienkatalog für Dienstleistungsgebäude – Neubau und Sanierung. Vienna. Austria, P. 31
34. Umweltbundesamt (n.d.). Ausschuss für Innenraumrichtwerte. Dessau-Roßlau. Germany https://www.umweltbundesamt.de/themen/gesundheit/kommissionen-arbeitsgruppen/ausschuss-fuer-innenraumrichtwerte#sitzungsprotokolle-des-air [Accessed 16-03-2023]
35. Federal Ministry for Climate Action, Environment, Energy, Mobility, Innovation and Technology (BMK) (2020). Richtlinie zur Bewertung der Luftqualität von Innenräumen https://www.bmk.gv.at/themen/klima_umwelt/luft/innenraum/rl_luftqualitaet.html [Accessed 16-03-2023]
36. Plesser. S. (2020). Qualitätsmanagement in der Gebäudetechnik. Technisches Monitoring und Inbetriebnahmemanagement. VDE Verlag GmbH. Berlin, Germany P. 21, P. 35
37. Zou, P. X. W., Wagle, D., & Alam, M. (2019). Strategies for minimizing building energy performance gaps between the design intend and the reality. Energy and Buildings, 191, 31-41. https://doi.org/10.1016/j.enbuild.2019.03.013
38. Pappalardo, M. & Reverdy, T. (2020). Explaining the performance gap in a French energy efficient building: Persistent misalignment between building design, space occupancy and operation practices. Energy Research & Social Science, 70, 101809. https://doi.org/10.1016/j.erss.2020.101809
39. Zou, P.X.W., Xu, X., Sanjayan, J., & Wang, J. (2018). Review of 10 years research on building energy performance gap: Life-cycle and stakeholder perspectives. Energy and Buildings, 178, 165-181. https://doi.org/10.1016/j.enbuild.2018.08.040.
40. Menezes, A.C., Cripps, A., Bouchlaghem, D., & Buswell, R. (2012). Predicted vs. actual energy performance of non-domestic buildings: Using post-occupancy evaluation data to reduce the performance gap. Applied Energy, 97, 355-364. https://doi.org/10.1016/j.apenergy.2011.11.075.
41. Vornanen-Winqvist, C., Järvi, K., Andersson, M. A., Duchaine, C., Létourneau, V., Kedves, O., Kredics, L., Mikkola, R., Kurnitski, J., & Salonen, H. (2020). Exposure to indoor air contaminants in school buildings with and without reported indoor air quality

problems. Environment International, 141, 105781. https://doi.org/10.1016/j.envint.2020.105781
42. Cai, C., Sun, Z., Weschler, L.B., Li, T., Xu, W., & Zhang, Y. (2021). Indoor air quality in schools in Beijing: Field tests, problems and recommendations. Building and Environment, 205, 108179. https://doi.org/10.1016/j.buildenv.2021.108179.
43. Singh, M.K., Ooka, R., Rijal, H.B., Kumar, S., Kumar, A., & Mahapatra, S. (2019). Progress in thermal comfort studies in classrooms over last 50 years and way forward. Energy and Buildings, 188-189, 149-174. https://doi.org/10.1016/j.enbuild.2019.01.051.
44. Teli, D., Bourikas, L., James, P.A.B., & Bahaj, A.S. (2017). Thermal Performance Evaluation of School Buildings using a Children-based Adaptive Comfort Model. Procedia Environmental Sciences, 38, 844-851. ISSN 1878-0296. https://doi.org/10.1016/j.proenv.2017.03.170.
45. Nam, I., Yang, J., Lee, D., Park, E., & Sohn, J.-R. (2015). A study on the thermal comfort and clothing insulation characteristics of preschool children in Korea. Building and Environment, 92, 724-733. https://doi.org/10.1016/j.buildenv.2015.05.041.
46. Rowland, T. (2008). Thermoregulation during exercise in the heat in children: old concepts revisited. Journal of Applied Physiology, 105(2), 718-724.
47. Al-Rashidi, K., Loveday, D., & Al-Mutawa, N. (2012). Impact of ventilation modes on carbon dioxide concentration levels in Kuwait classrooms. Energy and Buildings, 47, 540-549. https://doi.org/10.1016/j.enbuild.2011.12.030.
48. Theodosiou, T.G., & Ordoumpozanis, K.T. (2008). Energy, comfort and indoor air quality in nursery and elementary school buildings in the cold climatic zone of Greece. Energy and Buildings, 40(12), 2207-2214. https://doi.org/10.1016/j.enbuild.2008.06.011.
49. Zeiler, W., & Boxem, G. (2009). Effects of thermal activated building systems in schools on thermal comfort in winter. Building and Environment, 44(11), 2308-2317. https://doi.org/10.1016/j.buildenv.2009.05.005.
50. Zhang, X., Zhao, C., Zhang, T., Xie, J., Liu, J., & Zhang, N. (2023). Association of indoor temperature and air quality in classrooms based on field and intervention measurements. Building and Environment, 229, 109925. https://doi.org/10.1016/j.buildenv.2022.109925.
51. Siccardi, S., & Villa, V. (2023). Trends in Adopting BIM, IoT and DT for Facility Management: A Scientometric Analysis and Keyword Co-Occurrence Network Review. Buildings, 13(1), 15. https://doi.org/10.3390/buildings13010015
52. Tang, S., Shelden, D. R., Eastman, C. M., Pishdad-Bozorgi, P., & Gao, X. (2019). A review of building information modeling (BIM) and the internet of things (IoT) devices integration: Present status and future trends. Automation in Construction, 101, 127-139. https://doi.org/10.1016/j.autcon.2019.01.020
53. Schweiger, G., Eckerstorfer, L. V., Hafner, I., Fleischhacker, A., Radl, J., Glock, B., Wastian, M., Rößler, M., Lettner, G., Popper, N., & Corcoran, K. (2020). Active consumer participation in smart energy systems. Energy and Buildings, 227, 110359. https://doi.org/10.1016/j.enbuild.2020.110359. ISSN 0378-7788.
54. Schranz., T., Corcoran, K., Schwengler, T., Eckersdorfer, L. & Schweiger, G. (2020). Mobile Application for Active Consumer Participation in Building Energy Systems. BauSIM 2020 – 8th Conference of IBPSA Germany and Austria. Conference Paper. Graz. Austria
55. Schober, P., Boer, C., & Schwarte, L. A. (2018). Correlation coefficients: Appropriate use and interpretation. Anesthesia & Analgesia, 126(5), 1763-1768.

56. Hedderich, J. & Sachs, L. (2018): Angewandte Statistik – Methodensammlung mit R. Springer Verlag. Berlin. Germany. P.129-140. P. 450-455
57. Austrian Labour Inspectorate (n.d.). Kommentierte Arbeitsstättenverordnung. Retrieved from https://www.arbeitsinspektion.gv.at/Arbeitsstaetten-_Arbeitsplaetze/Arbeitsstaetten-_Arbeitsplaetze_1/Kommentierte_Arbeitsstaettenverordnung.html#heading___1_Anwendungsbereich [Accessed 31-03-2023]
58. Rowe, B. R., Canosa, A., Meslem, A., & Rowe, F. (2022). Increased airborne transmission of COVID-19 with new variants, implications for health policies. Building and Environment, 219, 109132. https://doi.org/10.1016/j.buildenv.2022.109132.
59. Federation of European Heating, Ventilation and Air Conditioning Association (n.d.). COVID-19 ventilation and building services guidance for school personnel, COVID-19 Task Force of REHVA's Technology and Research Committee. Brussels. Belgium
60. Centers for Disease Control and Prevention – CDC (2021). Ventilation in Buildings Retrieved from https://www.cdc.gov/coronavirus/2019-ncov/community/ventilation.html [Accessed 31-03-2023]
61. Ma, X., Liu, H., Zhang, Z., & Li, Y. (2023). How does indoor physical environment differentially affect learning performance in various classroom types? Building and Environment, 234, 110189. https://doi.org/10.1016/j.buildenv.2023.110189. ISSN 0360-1323.
62. Lamberti, G., Salvadori, G., Leccese, F., Fantozzi, F., & Bluyssen, P. M. (2021). Advancement on thermal comfort in educational buildings: Current issues and way forward. Sustainability, 13(18), 10315. https://doi.org/10.3390/su131810315.
63. Mudenda, S., Mukosha, M., Mwila, C., Saleem, Z., Kalungia, A. C., Munkombwe, D., Daka, V., Witika, B. A., Kampamba, M., Hikaambo, C. N., Sadiq, M. J., Chileshe, M., Kasanga, M., Mufwambi, W., Mfune, R. L., Matafwali, S. K., Masebe, P. O., Muungo, L. T., Bwalya, A. G., Kampamba, R. M., Zingani, E., Banda, D. C., Sintema, E. J., Gupta, A., Abdulrahman, N. M., Hangoma, J. M., Phiri, M. N., Hang'andu, D., Ngazimbi, M., Mudenda, F., Banda, M., & Kazonga, E. (2021). Impact of the coronavirus disease on the mental health and physical activity of pharmacy students at the University of Zambia: a cross-sectional study. International Journal of Basic & Clinical Pharmacology, 10(4), 324–332. https://doi.org/10.18203/2319-2003.ijbcp20211010
64. Hviid, C. A., Pedersen, C., & Dabelsteen, K. H. (2020). A field study of the individual and combined effect of ventilation rate and lighting conditions on pupils' performance. Building and Environment, 171, [106608]. https://doi.org/10.1016/j.buildenv.2019.106608
65. Baldinger, S., Baumann, R., Damberger, B., Hanus-Illnar, A., Hutter, H., Kundi, M., Palmsisano, G., Püringer, J., Schneider, J. Tappler, P.,(n.d.): Bewertung der Innenraumluft – Allgemeiner Teil, Working Group on Indoor Air Quality at the Federal Ministry of Agriculture, Forestry, Environment and Water Management (BMLFUW) Vienna, Austria. P. 12
66. Wang, D., Song, C., Wang, Y., Xu, Y., Liu, Y., & Liu, J. (2020). Experimental investigation of the potential influence of indoor air velocity on students' learning performance in summer conditions. Energy and Buildings, 219, 110015. https://doi.org/10.1016/j.enbuild.2020.110015. ISSN 0378-7788.
67. Halhoul Merabet, G., Essaaidi, M., Ben Haddou, M., Qolomany, B., Qadir, J., Anan, M., Al-Fuqaha, A., Abid, M. R., & Benhaddou, D. (2021). Intelligent building control systems for thermal comfort and energy-efficiency: A systematic review of artificial intelligence-assisted techniques. Renewable and Sustainable Energy Reviews, 144, 110969. https://doi.org/10.1016/j.rser.2021.110969. ISSN 1364-0321.

1. Quintana, M., Schiavon, S., Tartarini, F., Kim, J., & Miller, C. (2023). Cohort comfort models — Using occupant's similarity to predict personal thermal preference with less data. Building and Environment, 227(Part 1), 109685. https://doi.org/10.1016/j.buildenv.2022.109685. ISSN 0360-1323.